\documentclass[twocolumn,showpacs,preprintnumbers,amsmath,amssymb,amsfonts,pra]{revtex4-1}
\usepackage{graphicx,epsfig}
\usepackage{dcolumn}
\usepackage{bm}

\begin{document}

\title{Asymmetric light transmission through a photonic crystal with relaxing Kerr nonlinearity}

\author{Denis V. Novitsky}
\email{dvnovitsky@tut.by} \affiliation{B.I. Stepanov Institute of
Physics, National Academy of Sciences of Belarus,
Nezavisimosti~Avenue~68, 220072 Minsk, Belarus}

\date{\today}

\begin{abstract}
The idea of asymmetric light transmission through a photonic crystal
with relaxing Kerr nonlinearity is presented. This idea is based on
the symmetry breaking due to the self-trapping of intensive pump
pulse inside the structure. As a result, the asymmetric transmission
of the secondary (probe) pulses through the perturbed photonic
crystal can be observed. The cases of short and long
(quasi-continuous) probe pulses are considered.
\end{abstract}


\maketitle

\section{Introduction}

The problem of asymmetric light propagation attracts much attention
in modern scientific literature. The essence of this phenomenon is
the situation when light transmission through the optical system in
two opposite directions give different radiation characteristics. In
the limiting case, one can observe the so-called optical diode
action when light is transmitted or entirely blocked depending on
the propagation direction. The importance of such effects is
connected with the perspectives of integrated optics and all-optical
logics and can lead to construction of all-optical diodes, isolators
and other devices. To achieve asymmetric or unidirectional
transmission, a number of optical structures were recently proposed
on the basis of different operation principles. A number of
suggestions is connected with the use of multilayer systems (or
one-dimensional photonic crystals), for example, the passive optical
diode based on nonlinear effects near the band edge~\cite{Scalora,
Tocci}, the hybrid Fabry-Perot-resonator–photonic-crystal structure
with strongly asymmetric resonant modes~\cite{Zhukovsky}, photonic
crystal heterostructure with broken spatial inversion
symmetry~\cite{Cuicui}, metal-dielectric multilayer diode due to a
light-tunneling mechanism~\cite{Chunhua}, asymmetric multilayer with
left-handed materials~\cite{Feise} and so on. A class of systems
utilizes the effects of photonic crystal defects to achieve
unidirectionality, such as coupling of nonlinear defects~\cite{Cai}
or microcavities with the modes of opposite parity~\cite{Grigoriev},
asymmetrically placed nonlinear defect in the structure with
chiral~\cite{Tuz1} or magnetic layers~\cite{Tuz2}, photonic crystal
waveguide with uncoupled microcavities which is subject to a
Fano-like mechanism and asymmetric light localization~\cite{Ding}.
In ref.~\cite{Feng} two-dimensional photonic crystal heterostructure
was used to obtain diode action due to a self-collimation behavior.
Another type of structures considered in literature uses the
so-called metamaterials, for example, planar split-ring resonators
with chiral symmetry breaking~\cite{Plum}, bilayered chiral
metamaterial with polarization-dependent asymmetry~\cite{Huang},
artificial structure with polarization selectivity due to
magnetoelectric coupling and light tunneling~\cite{Mutlu}. Asymmetry
in transmission was recorded in the nonlinear waveguide with a
localized phase discontinuity~\cite{Gallo}, nonlinear directional
couplers with longitudinally varying coupling
coefficient~\cite{Tran}, etc. A similar effect of asymmetric
reflection suppression (known as unidirectional invisibility) was
found in parity-time symmetric periodic structures~\cite{Lin, Ge}.
Finally, it is worth to mention the possibility of asymmetric
transmission of ultrashort pulses through a resonantly absorbing
two-level medium discussed in ref.~\cite{Novit12}.

In this paper we theoretically search for evidence of asymmetric
transmission of ultrashort light pulses interacting with a
one-dimensional photonic crystal (multilayer structure) made from
materials possessing relaxing cubic (Kerr) nonlinearity. The need to
take noninstantaneousness of nonlinearity into account was realized
soon after appearance of nonlinear optics as a separate discipline.
To name a few examples, one can mention studies of relaxation
effects on laser beam self-focusing~\cite{Fleck} or parametric
amplification~\cite{Trillo}. The effects of finite relaxation time
become apparent when this time is comparable with pulse duration.
This situation seems to be very actual nowadays as the methods of
production of ultrashort (femtosecond) pulses are established. In
the context of photonic crystals, the concept of relaxing Kerr
nonlinearity was applied to the phenomena such as pulse compression
\cite{Vlasov} and optical switching \cite{Meng}. The latter work
also suggests that the polymeric materials are perspective media for
applications due to both high nonlinearity coefficient and fast
relaxation.

The main feature of the nonlinear photonic structure important for
our study is the self-trapping effect analyzed in our previous
works~\cite{Novit, Novit1, Novit2}. This effect occurs when both
periodic modulation of linear refractive index and relaxation of
nonlinearity are present. In this letter we suggest that
self-trapping can be used to break the symmetry of the photonic
crystal: the powerful pump pulse trapped inside the system changes
the distribution of the refractive index which, in turn, influences
the interaction of the secondary (probe) pulse with the structure.
Such a variant of pump-probe technique is possible due to the
relaxation of nonlinearity and allows to use pump and probe one
after another. Simultaneously, this solves the common problem of
separation between pump and probe signals. The paper includes the
formulation of our theoretical model, the description of the main
features of self-trapping effect, the results on asymmetric
transmission in the cases of short-pulse and quasi-continuous
regimes. Finally, we discuss the importance of probe intensity to
obtain the asymmetric transmission.

\section{The description of the model}

\begin{figure}[t!]
\includegraphics[scale=0.9, clip=]{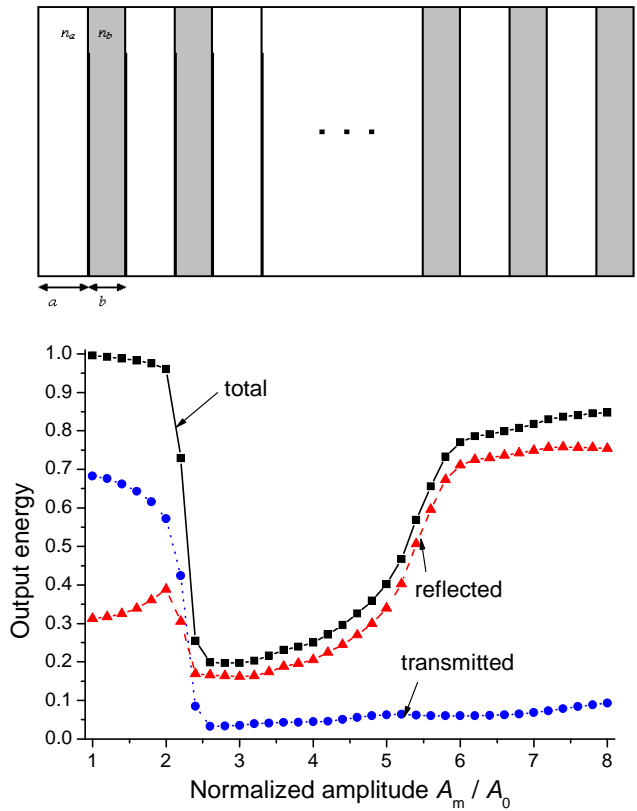}
\caption{\label{fig1} (Upper panel) The scheme of the
one-dimensional photonic crystal. (Lower panel) Dependence of the
output light energy (normalized to the input energy) on the peak
amplitude of the incident pulse. The energy is integrated over the
time $300 t_p$.}
\end{figure}

In this letter we consider an ultrashort (femtosecond) pulse
interaction with nonlinear photonic crystal taking into account the
relaxation process. Light propagation in this system is governed by
the Maxwell wave equation,
\begin{eqnarray}
\frac{\partial^2 E}{\partial z^2}&-&\frac{1}{c^2} \frac{\partial^2
(n^2 E)}{\partial t^2} = 0, \label{Max}
\end{eqnarray}
where $E$ is the electric field strength, $c$ is the light speed,
$n=n(z, t)$ is the nonlinear refractive index. The relaxation of
nonlinearity is described by the Debye model,
\begin{eqnarray}
t_{nl} \frac{d \delta n (I, t)}{d t}+ \delta n (I, t)=n_2 I,
\label{relax}
\end{eqnarray}
where $\delta n$ is the nonlinear part of refractive index, $n_2$ is
the Kerr nonlinear coefficient, $I=|E|^2$ is the light intensity.
The full refractive index is
\begin{eqnarray}
n(z, t)=n_0(z)+\delta n (I, t), \label{refr}
\end{eqnarray}
where $n_0(z)$ is the linear part of the refractive index varying
periodically along the $z$ axis and defining the structure of the
photonic crystal. To simulate pulse propagation in this system, we
solve self-consistently together with the eqs.~(\ref{relax}) and
(\ref{refr}) using the finite-difference time-domain method
described in ref.~\cite{Novit}. In our calculations we assume the
light pulses to have the Gaussian shape with the amplitude $A=A_m
\exp(-t^2/2t_p^2)$, the duration $t_p=30$ fs, and the central
wavelength of the initial pulse spectrum $\lambda_c=1.064$ $\mu$m.
As regards the photonic crystal (see the upper panel of fig.
\ref{fig1}), the linear parts of refractive indices of the
alternating layers are $n_a=2$ and $n_b=1.5$, respectively; their
thicknesses $a=0.4$ and $b=0.24$ $\mu$m; the number of layers
$N=200$. The nonlinear coefficient of the materials is defined
through the nonlinear term of the refractive index, so that $n_2
I_0=0.005$; this means that the pulse amplitude is normalized by the
value $A_0=\sqrt{I_0}$. The relaxation time of the nonlinearity of
all layers is $t_{nl}=10$ fs (fast electronic Kerr nonlinearity).
Note that we adopt equal nonlinearity parameters for both types of
the layers that do not limit generality as was shown in the study of
the different variants of the structure including the half-linear
one~\cite{Novit2}. The changes in light spectra in such situation
were studied in ref.~\cite{Novit1}, so that we do not consider the
spectral effects in this paper.

\section{Self-trapping effect and trap position}

The essence of self-trapping inside the photonic crystal is the
localization of light pulse due to formation of a nonlinear "cavity"
(or trap) which dynamically adjusts its properties to light
intensity and which is stabilized by nonlinearity relaxation. This
trap has a form of bell-shaped distribution of intensity and
refractive index variation. This variation preserves its form for a
relatively long time (more than $100$ transmission times in the
linear case)~\cite{Novit}. Thus, there are three temporal parameters
of the system: pulse duration $t_p$, nonlinearity relaxation time
$t_{nl}$, and pulse propagation time $t_{prop}$. Obviously, the
effects of relaxation become important when $t_p$ and $t_{nl}$ are
comparable; $t_{prop}$ must be not too short for relaxation
processes to have enough time to show themselves. In other words,
the structure must be long enough for a pulse to be trapped inside
it. For our photonic crystal described above, $t_{prop} \sim 30 t_p$
which is a large enough value.

The effect of self-trapping can be indicated by the value of light
energy leaving the photonic crystal as a result of reflection and
transmission. This value is calculated by the integration of
radiation intensity over the certain time interval. If this output
energy normalized by the energy of the initial pulse and integrated
over a large enough time is much less than the unity, one can claim
that the pulse is trapped inside the structure. The results of
calculation of the output energy as a function of the incident pulse
amplitude for the above stated parameters are shown in
fig.~\ref{fig1}. This value is integrated over the time $300 t_p$
which is about $10$ times larger than the propagation time of the
pulse through the photonic crystal in linear regime. It is seen that
for the amplitudes larger than the critical one (about $2.2 A_0$),
the output energy abruptly drops which is the evidence of the pulse
self-trapping. There is the region of optimal light-matter coupling.
However, as the amplitude increases, the output energy goes up due
to reflection. This is explained by the trapping of the pulse near
the very entrance of the structure~\cite{Novit}.

\begin{figure}[t!]
\includegraphics[scale=0.9, clip=]{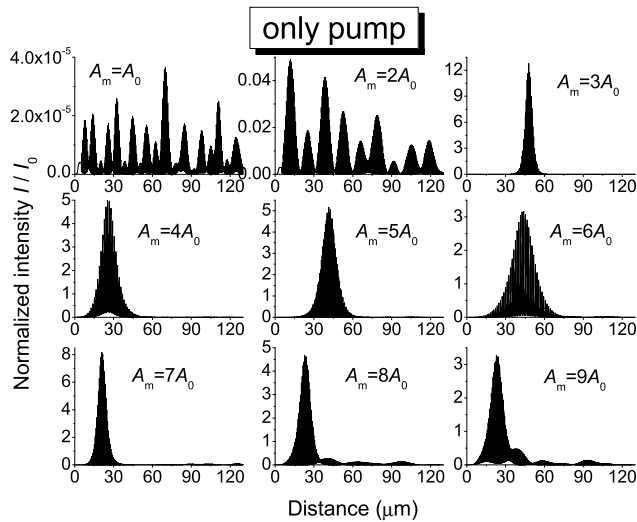}
\caption{\label{fig2} Distribution of light intensity inside the
photonic crystal (at $t=200 t_p$) for different pump pulse
amplitudes.}
\end{figure}

As it was mentioned above, the result of the pulse self-trapping is
the formation of the quasi-stationary trap confining most part of
the pulse energy. This trap can be viewed as a characteristic
distribution of light intensity inside the system (its total length
is $128$ $\mu$m). In fig.~\ref{fig2} we plot such distributions at
the instant $t=200 t_p$ after the pulse incidence and at different
values of the pulse amplitude. It is seen that at low amplitudes
($A_m=A_0$ and $2 A_0$), the intensity of radiation in the photonic
crystal is very low and its distribution looks disordered and is
extended over the whole structure. Above the critical amplitude
($A_m=3 A_0$ and so on), the distribution takes a bell-shaped form
(due to the similar shape of the pulse) which is formed closer to
the entrance of the system for stronger pulse amplitudes. This
change of position is connected with the increase of reflection with
the amplitude noted in fig.~\ref{fig1}. Simultaneously, the peak
value of intensity inside this trap tends to decrease (the trap gets
wider) though its absolute value can sometimes become larger due to
a larger absolute value of the pulse amplitude.

The most important feature of the distributions considered is their
position inside the photonic crystal. The initial pulse propagates
from left to right, so that the position of the trap is closer or
farther from the left side of the structure depending on the pulse
amplitude. Obviously, the photonic crystal which was initially
perfectly symmetric obtains some attribute of asymmetry when the
trap is located inside it. Let us calculate how this asymmetry shows
itself in the secondary (probe) pulse propagation in two variants:
from left to right (LR) and vice versa (RL). The initial pulse
forming the trap is further called the pump pulse and is launched at
the instant $t=0$ (the time of maximum incidence at the input).

\section{Short probe pulse transmission}

\begin{figure}[t!]
\includegraphics[scale=0.9, clip=]{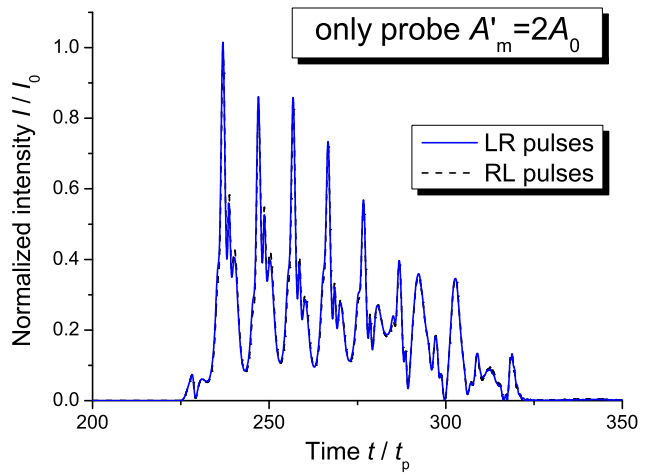}
\caption{\label{fig3} Intensity profiles of $10$ probe pulses
transmitted from left to right (LR) or vice versa (RL) in absence of
the pump pulse. The amplitude of the probe pulses is $A'_m=2 A_0$,
the interval between the pulses is $10 t_p$; the first pulse starts
at $t=200 t_p$.}
\end{figure}

To study the problem of asymmetry, we take a series of $10$ probe
pulses: the first starts at the instant $t=200 t_p$, while the
interval between the other is $10 t_p$. All probe pulses are
identical and have the same subcritical amplitude $A'_m=2 A_0$.
First of all, we consider propagation of LR and RL pulses in the
absence of any pump pulse, i.e. without a trap inside the structure.
Figure~\ref{fig3} shows the resulting profiles of $10$ subcritical
probe pulses transmitted through the nonlinear photonic crystal. One
can easily see that the transmission in this case is perfectly
symmetric: the profiles are identical for both directions, though
every individual pulse behaves in a different way comparing with its
neighbors. This is the evidence of interaction between the pulses
which occurs due to the shortage of the interpulse interval.

\begin{figure}[t!]
\includegraphics[scale=0.9, clip=]{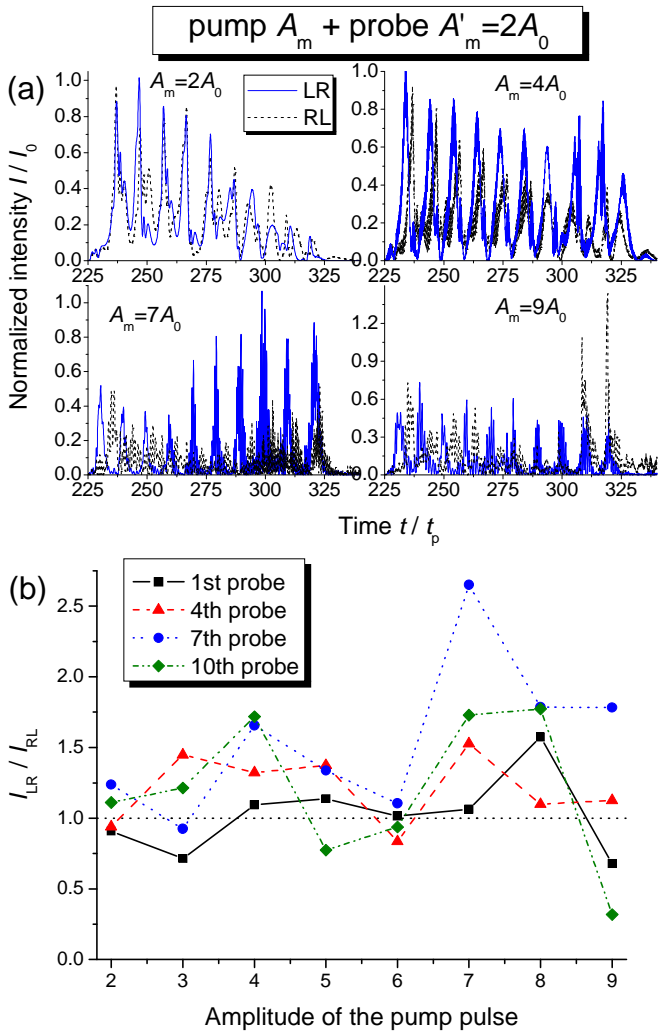}
\caption{\label{fig4} (a) Intensity profiles of $10$ probe pulses
transmitted from left to right (LR) or vice versa (RL) for different
amplitudes $A_m$ of the pump pulse forming the trap. (b) The
relation of the peak intensities $I_{LR}/I_{RL}$ of the 1st, 4th,
7th, and 10th probe pulses versus the amplitude of the pump. The
parameters of the probe pulses are the same as in fig.~\ref{fig3}.}
\end{figure}

Now we can analyze the influence of light distributions formed by
the initial (pump) pulse (see fig.~\ref{fig2}) on the properties of
the transmitted probe pulses. The results of calculations of
intensity profiles at different amplitudes of the pump pulse are
depicted in fig.~\ref{fig4}(a). The probe pulses start at $t=200
t_p$ as previously. It is seen that even at the subcritical
amplitude of the pump pulse ($A_m=2A_0$), when the residual light
intensity inside the photonic crystal is very low ($I \sim 10^{-6}
I_0$), the perfect symmetry between LR and RL transmission is
already broken. Of course, the difference between these two cases is
not dramatic, though it gets more pronounced for the latest probe
pulses than for the first ones. This is the case for all the
variants shown in fig.~\ref{fig4}(a) and is perhaps connected with
the influence of the first pulses on the trap. However, the
correspondence between the LR and RL pulses is still close for
$A_m=4A_0$. For larger amplitudes ($A_m=7A_0$ and $9A_0$), as the
trap forms closer to the entrance of the system, the difference
appears even between the first probe pulses moving in opposite
directions. They are not equal not only in intensity, but also they
need significantly different times to transmit through the
structure: LR pulses propagate faster than RL ones. The disparity in
intensity grows for latest pulses, so that they can differ by a
factor of two and more. This point is illustrated by
fig.~\ref{fig4}(b) where the ratio of the peak intensities
$I_{LR}/I_{RL}$ is shown. It is seen that the deviation of this
value from unity tends to get greater for the latest probe pulses
and for larger pump amplitudes. This figure also suggests the choice
of proper pump intensity to achieve the optimal contrast between the
LR and RL transmission.

\section{Quasi-continuous regime}

\begin{figure}[t!]
\includegraphics[scale=0.9, clip=]{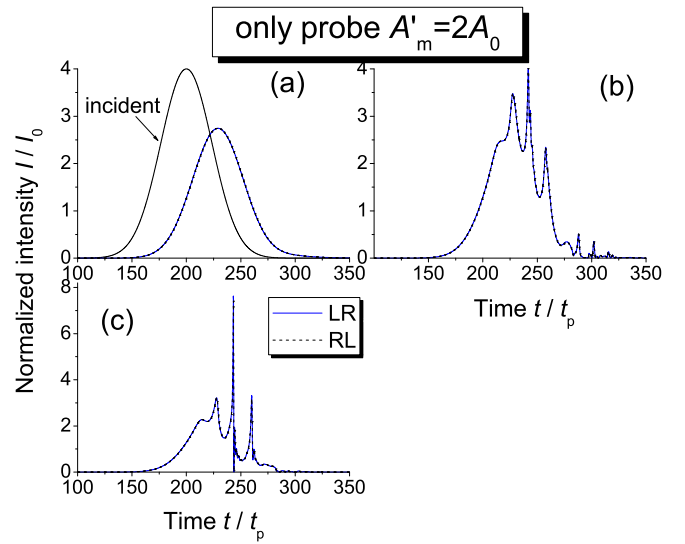}
\caption{\label{fig5} Intensity profiles the quasi-continuous probe
pulse transmitted from left to right (LR) or vice versa (RL) in
absence of the pump pulse. (a) The case of the linear photonic
crystal; (b) nonlinearity without relaxation; (c) relaxing
nonlinearity. The amplitude of the probe pulse is $A'_m=2 A_0$, its
duration is $1$ ps; the pulse starts at $t=200 t_p$ (as previously,
$t_p=30$ fs).}
\end{figure}

In the previous section, we have considered the case of probe pulses
of the same duration as the pump (trapped) pulse. Such situation
when the probe pulses effectively experience the relaxation of
nonlinearity (due to $t_p \sim t_{nl}$) should be discriminated from
the quasi-continuous regime. In this last case $t_p \gg t_{nl}$, so
that the influence of the relaxation on the pulse propagation should
be not so dramatic. Let us analyze propagation of the probe pulse
with the duration $t'_p=1$ ps, while leaving the relaxation time
($t_{nl}=10$ fs) and pump pulse duration ($t_p=30$ fs) unchanged.
The results of calculations in absence of pump are plotted in
fig.~\ref{fig5} for different cases: the linear photonic crystal,
the nonlinear one without relaxation, and, finally, the nonlinear
structure with relaxing Kerr nonlinearity. As previously, we see the
symmetry in pulse transmission in all cases. It is also worth to
point to the narrow peaks appearing in the profiles of the pulse
transmitted through the system with nonlinearity. Obviously, this is
the clear evidence of the well-known effect of modulation
instability which leads to the breakup of the long pulse into a
train of short ones~\cite{Agrawal}. As the comparison between
figs.~\ref{fig5}(b) and \ref{fig5}(c) shows, the relaxation of
nonlinearity promotes the modulation instability and precipitates
the breakup. The study of these processes is beyond the scope of
this paper, therefore we only make here this short comment and turn
to discussion of the influence of the pump pulse on the probe
behavior.

\begin{figure}[t!]
\includegraphics[scale=0.85, clip=]{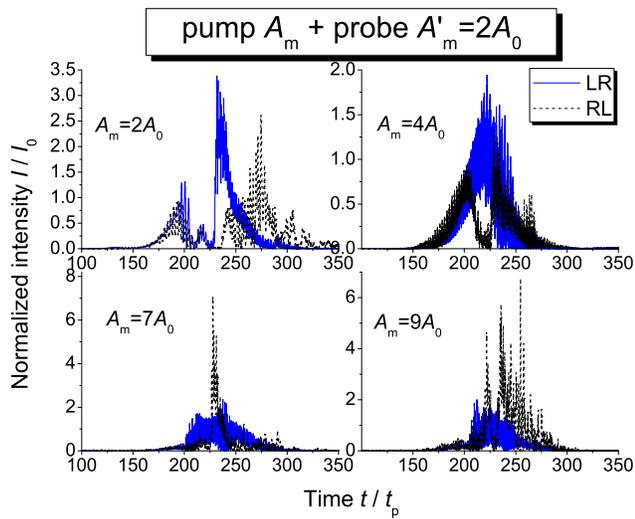}
\caption{\label{fig6} Intensity profiles of the quasi-continuous
probe pulse transmitted from left to right (LR) or vice versa (RL)
for different amplitudes $A_m$ of the pump pulse forming the trap.
The parameters of the probe pulses are the same as in
fig.~\ref{fig5}. The duration of the pump pulse is still $t_p=30$
fs.}
\end{figure}

The dramatically increased efficiency of this influence can be
easily seen in fig.~\ref{fig6}. Even a subcritical pump pulse
($A_m=2 A_0$) appears to be enough to break the symmetry in the
propagation of the quasi-continuous probe pulse. Perhaps, this is
connected with the heightened sensitivity of such nonlinear
processes as the modulation instability to the changes in external
conditions. This results in the strong change of the pulse profile,
which, together with the asymmetric distribution of the refractive
index, leads to the strengthening of the dependence of transmission
on the propagation direction. As a result, the sharp rise-up portion
of the LR pulse is obtained with the subsequent gradual decrease in
intensity. The profile of the RL pulse is fundamentally different.
However, at larger amplitudes of the pump pulse ($A_m=7 A_0$ and $9
A_0$), the situation is inverted: the sharp peaks for the RL pulse
and unstructured radiation for the LR one. This can be explained as
a result of the probe pulse interaction with the trap localized near
the very left end of the photonic crystal: under these conditions,
the LR probe is perturbed in the very beginning of propagation and
does not have enough time to form a characteristic profile. The
contrast in light intensity also increases at higher amplitudes of
the pump, so that the peak intensity in one direction exceeds the
same value in the other by the factor of three as can be seen in the
lower panels of fig.~\ref{fig6}.

\section{Is the intensity of probe important?}

\begin{figure}
\includegraphics[scale=0.81, clip=]{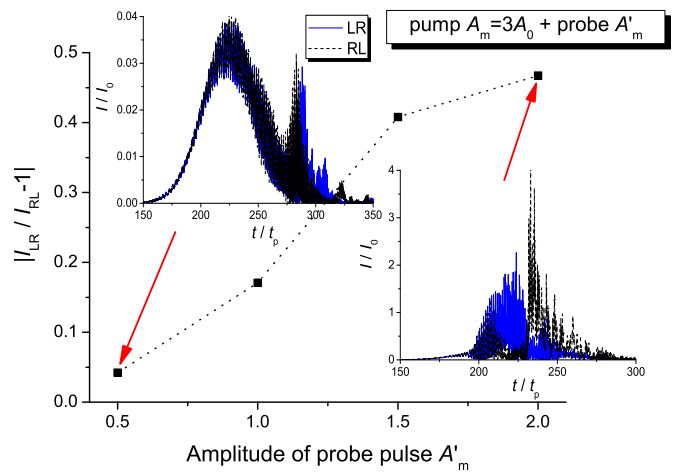}
\caption{\label{fig7} The value $|I_{LR}/I_{RL}-1|$ for the
quasi-continuous probe pulse transmitted from left to right (LR) or
vice versa (RL) versus the amplitude of the probe. The insets show
the probe pulse profiles for the cases of relatively large and low
probe intensities. The pump pulse has amplitude $A_m=3A_0$ and
duration $t_p=30$ fs. The other parameters are the same as in
fig.~\ref{fig5}.}
\end{figure}

This question should by studied to understand, whether the
self-trapping effect is the only process responsible for asymmetric
propagation or the interaction of the probe pulse with nonlinear
structure must be taken into account as well. Previously we have
considered probe pulses with relatively strong amplitude ($2 A_0$),
not far from the amplitude of self-trapping. Since we are interested
only in absolute value of asymmetry, we calculate the quantity
$|I_{LR}/I_{RL}-1|$ as a characteristic of direction dependence. The
results of calculations for the quasi-continuous $1$-ps-long probe
pulse are shown in fig.~\ref{fig7}. One can see that the value of
asymmetry strongly depends on the probe amplitude and is close to
zero already for $A'_m=0.5 A_0$. The insets visually demonstrate the
difference in pulse profiles in the cases of strongly asymmetrical
and almost symmetrical propagation. Similar dependence takes place
for short probe pulses (not shown here).

The results of fig.~\ref{fig7} imply that the strong interaction of
the probe pulse with the nonlinear photonic structure is necessary
to observe asymmetric transmission as well as additional refractive
index modulation due to self-trapping. In other words, strong
amplitudes of both pump and probe are needed to obtain the described
effect. It is hard to demarcate the influence of pump self-trapping
and probe interaction in the situations considered. Perhaps, the
study of change of asymmetry as a function of slow temporal change
of the trap can give some additional clues. However, the situation
seems to be similar to that described in ref.~\cite{Zhukovsky} where
the asymmetric distribution of refractive index is used to localize
the radiation which further nonlinearly interacts with the
structure. In our case, the asymmetry in refractive index is caused
by the pump pulse (quasi-stationary trap), then the probe pulse
nonlinearly interacts with the modified periodic system. We should
also emphasize that the nonlinearity relaxation is necessary to
obtain the trap \cite{Novit} and, hence, to obtain a
quasi-stationary (and large) change of materials refractive indices
which influence the probe pulse. It is also worth noting that high
probe intensity makes the proposed scenario of asymmetric
transmission different from a typical pump-probe technique.

\section{Conclusion}

To sum it up, we have considered the possibility of asymmetric light
transmission in the regime of pulse self-trapping in the photonic
crystal with relaxing Kerr nonlinearity. It is shown that the
position of the trap formed by the high-intensity pump pulse
strongly influences the propagation of probe pulses and leads to the
dependence of the transmission characteristics on the propagation
direction. In the case of a series of short probe pulses, this
asymmetry is sharp enough only for the late components of the
series, while the single long (quasi-continuous) pulse feels the
pump radiation much stronger. The light dynamics in this latter case
are more complex and include the processes such as the modulation
instability. As a result, the efficiency of asymmetric transmission
is higher under the quasi-continuous conditions than in the short
pulse regime. The dependence of this effect on probe pulse intensity
allows to suggest that the role of pump comes to formation of
asymmetric refractive index modulation which modifies nonlinear
interaction of probe with the material of photonic crystal.
Additional evidence for the asymmetric transmission is likely to be
given in the study of the effects such as asymmetric bistable
response.

\textbf{Acknowledgement.} The work was supported by the Belarusian
Foundation for Fundamental Research (Grant No. F11M-008).

\end{document}